\documentclass[conference]{IEEEtran}
\IEEEoverridecommandlockouts
\usepackage{cite}
\usepackage{amsmath,amssymb,amsfonts}
\usepackage{algorithmic}
\usepackage{algorithm}
\usepackage{graphicx}
\usepackage{textcomp}
\usepackage{xcolor}
\usepackage{booktabs}
\usepackage{multirow}
\usepackage{tikz}
\usetikzlibrary{shapes,arrows,positioning}
\usepackage{url}
\usepackage{hyperref}
\usepackage{cleveref}
\usepackage{orcidlink}

\begin{document}

\title{An Empirical Evaluation of Quantum-Inspired QUBO Methods for Heterogeneous HPC Workflow Mapping and Scheduling%
\thanks{To appear in: \textit{Proc. 41st Int. Conf. ISC High Performance 2026 (ISC26)}, Hamburg, Germany, June 22 to 26, 2026, IEEE Xplore. Published open access under Creative Commons Attribution 4.0 (CC BY 4.0). This is the authors' camera-ready version. \copyright~2026 The Authors.}}

\author{
\IEEEauthorblockN{Aasish Kumar Sharma\,\orcidlink{0000-0002-7514-2340}\IEEEauthorrefmark{1}, Christian Boehme\,\orcidlink{0000-0002-4289-0465}\IEEEauthorrefmark{2}, Julian Kunkel\,\orcidlink{0000-0002-6915-1179}\IEEEauthorrefmark{1}\IEEEauthorrefmark{2}}
\IEEEauthorblockA{\IEEEauthorrefmark{1}Institute of Computer Science, Georg-August-Universit\"at G\"ottingen, Germany}
\IEEEauthorblockA{\IEEEauthorrefmark{2}GWDG mbH, G\"ottingen, Germany}
\IEEEauthorblockA{Email: \{christian.boehme, julian.kunkel, aasish-kumar.sharma\}@gwdg.de}
}

\maketitle

\begin{abstract}
Heterogeneous HPC workflow scheduling under multiple hard constraints poses a challenging combinatorial optimization problem. Classical exact solvers provide optimal solutions but often face scalability limitations, motivating interest in quantum-inspired Quadratic Unconstrained Binary Optimization (QUBO) formulations as alternative optimization paradigms. This work presents a systematic and reproducible empirical evaluation of QUBO-based scheduling methods against established classical baselines, including MILP, CP-SAT, GA, and HEFT. We evaluate three QUBO variants, single-run simulated annealing, multi-attempt annealing, and a layered QAOA-inspired schedule, together with hybrid enhancement strategies on both ground-truth validation workflows (3-4 tasks) and synthetic scaling instances (5-20 tasks). All solvers are assessed through a unified evaluation pipeline that explicitly tracks feasibility, makespan, and resource utilization under progressive constraint activation and controlled penalty sweeps. Results show that all approaches recover the expected optimal makespan on validation instances, confirming formulation correctness. However, feasibility degradation emerges for specific QUBO variants as constraint interactions intensify, particularly when communication costs are introduced. Penalty sensitivity analysis reveals a sharp feasibility threshold for QUBO-SA, where insufficient penalties consistently fail and moderate-to-strong penalties restore feasibility. Scaling experiments delineate the regimes in which classical solvers and heuristics remain robust across all tested sizes, while QUBO-SA loses feasibility beyond 15 tasks and the QAOA-inspired variant beyond 10 tasks. Overall, the study provides a clear empirical characterization of the reliability boundaries of quantum-inspired QUBO formulations for heterogeneous HPC scheduling and identifies the regimes where classical approaches remain preferable under current formulations and solver capabilities.
\end{abstract}

\begin{IEEEkeywords}
Workflow mapping and scheduling, Mixed Integer Linear Programming (MILP), QUBO formulation, simulated annealing, quantum computing, constraint programming, Heterogeneous Earliest Time First (HEFT), HPC systems
\end{IEEEkeywords}

\section{Introduction}

Workflow mapping and scheduling in heterogeneous high-performance computing (HPC) systems requires assigning interdependent tasks to diverse compute resources while satisfying precedence, capacity, and feature constraints.
This problem is inherently combinatorial and grows rapidly in complexity with increasing workflow size and system heterogeneity.
Classical exact approaches such as mixed-integer linear programming (MILP) and constraint programming (CP-SAT) can guarantee optimality but face practical scalability limits, while heuristic methods such as HEFT offer efficient but non-optimal solutions \cite{topcuoglu2002performance,laborie2018ibm}.

Recent advances in quantum and quantum-inspired optimization have renewed interest in quadratic unconstrained binary optimization (QUBO) as an alternative formulation framework.
QUBO models are theoretically universal and can encode combinatorial problems using binary variables and quadratic interactions \cite{glover2018tutorial}.
Such formulations are compatible with quantum annealers, gate-based quantum algorithms such as QAOA, and classical solvers via simulated annealing \cite{farhi2014quantum,albash2018adiabatic}.

However, applying QUBO to realistic scheduling remains challenging.
Hard constraints are enforced via penalty terms whose calibration becomes difficult when multiple heterogeneous constraints interact.
Much of the literature evaluates QUBO or QAOA on carefully selected instances, with limited systematic comparison to state-of-the-art classical baselines or explicit characterization of feasibility and scalability limits.

This work provides a rigorous empirical assessment of quantum-inspired QUBO formulations for heterogeneous HPC mapping and scheduling, focusing on feasibility, penalty sensitivity, and scalability relative to established classical methods.
The motivation is not to claim QUBO superiority, but to establish clear empirical evidence on where and when penalty-based QUBO formulations succeed, fail, and what conditions govern the transition between these regimes.
Such evidence is essential for the HPC community to make informed decisions about adopting quantum-inspired methods and for guiding future quantum hardware evaluation.

\subsection{Research Questions}

We address the following research questions:

\textbf{RQ1:} Under which workload characteristics and constraint structures do QUBO-based methods become competitive with classical mapping and scheduling approaches?

\textbf{RQ2:} What trade-offs arise when mapping multi-constraint HPC scheduling problems into penalty-based QUBO formulations, particularly in feasibility and scalability compared to CP-SAT?

\textbf{RQ3:} What empirical regimes characterize the reliability limits of QUBO-based solvers for production-scale HPC workflows?

\textbf{RQ4:} Do QUBO-based approaches exhibit an operational advantage over classical baselines on this workload class under fixed computational budgets, measured by feasibility, makespan quality, and runtime?

Each research question maps to specific experiments and measurable metrics:
RQ1 is addressed through progressive constraint activation (Experiment~0), measuring feasibility under increasing constraint complexity.
RQ2 is addressed through penalty sensitivity sweeps (Experiment~1), measuring feasibility rate and makespan gap as a function of penalty multiplier.
RQ3 is addressed through scaling experiments (Experiment~3), measuring feasibility boundaries across 5 to 20 tasks.
RQ4 synthesizes evidence from all experiments, comparing QUBO methods against MILP, CP-SAT, GA, and HEFT on feasibility, makespan quality, and runtime under fixed computational budgets.

\subsection{Contributions}

Our main contributions are:
(i) A formally correct QUBO formulation for feature-aware task-node mapping in heterogeneous HPC systems.
(ii) A set of enhancement strategies addressing penalty calibration and feasibility, including adaptive penalties and hybrid classical repair.
(iii) A unified evaluation pipeline combining QUBO-based mapping with classical schedule derivation for makespan and utilization analysis.
(iv) A systematic experimental study spanning validation benchmarks and controlled scaling experiments (5-20 tasks), benchmarking QUBO methods against MILP, CP-SAT, GA, and HEFT.

\section{Background and Related Work}

\subsection{Workflow Scheduling Fundamentals}

HPC workflows are commonly modeled as directed acyclic graphs (DAGs), where nodes represent tasks and edges capture precedence and data dependencies.
Each task has resource and feature requirements, while compute nodes provide limited capacity and heterogeneous capabilities.
Scheduling seeks to minimize makespan and maximize resource utilization subject to five hard constraints: unique assignment, capacity limits, feature compatibility, precedence, and communication costs.

\subsection{Classical Scheduling Approaches}

MILP formulations model scheduling with binary assignment variables and linear constraints, offering provable optimality at the cost of exponential worst-case complexity \cite{drozdowski2009scheduling}.
Constraint programming, particularly CP-SAT, exploits domain propagation and global constraints to solve moderately sized scheduling problems efficiently \cite{laborie2018ibm}.
Heuristic approaches such as HEFT compute task priorities and greedily assign tasks, achieving near-optimal performance at low computational cost \cite{topcuoglu2002performance}.
Population-based methods like genetic algorithms provide broader search but incur higher runtime overhead \cite{yu2006scheduling}.

\subsection{Quantum-Inspired Optimization and QUBO}

QUBO expresses optimization problems as quadratic functions over binary variables and serves as the native input for quantum annealers and QAOA-based algorithms \cite{glover2018tutorial}.
Simulated annealing provides a classical approximation, while quantum hardware implementations remain limited in scale and noise tolerance \cite{king2018observation}.

Penalty-based constraint encoding is central to QUBO modeling but introduces a critical challenge: penalties must be large enough to enforce feasibility without overwhelming objective optimization.
Empirical studies consistently observe sharp feasibility thresholds rather than smooth trade-offs, complicating manual tuning \cite{glover2018tutorial}.

\subsection{QAOA, Multi-Objective Optimization, and Limitations}

QAOA was introduced as a variational quantum algorithm for approximate optimization \cite{farhi2014quantum} and later generalized to the Quantum Alternating Operator Ansatz to better handle constraints \cite{hadfield2019quantum}.
Despite extensive study, no general quantum advantage for QAOA has been established; theoretical results show fundamental limitations for constant-depth circuits \cite{boulebnane2024solving}.

Recent work reports empirical scaling advantages for QAOA on highly structured problems such as LABS under idealized assumptions \cite{shaydulin2024evidence}, and extensions to multi-objective optimization have demonstrated Pareto-front approximation capabilities \cite{kotil2025quantum}.
However, these results rely on problem structures and assumptions that differ substantially from realistic HPC mapping and scheduling, which involves multiple interacting hard constraints and heterogeneous resources.

Consequently, recent survey and thesis work emphasize hybrid quantum-classical approaches, where QUBO-based mappings are combined with classical feasibility repair or scheduling decoders, as the most practical near-term pathway \cite{Mete2023}.

\subsection{Research Gap}

Existing studies largely evaluate QUBO or QAOA methods in isolation, without rigorous comparison to classical exact and heuristic solvers across increasing constraint complexity.
There is a lack of systematic evidence identifying when penalty-based QUBO formulations remain reliable and when classical methods dominate.

This work addresses this gap through controlled benchmarking, progressive constraint activation, and scaling experiments, providing an empirical foundation for assessing the realistic role of quantum-inspired optimization in HPC scheduling.

\section{Methodology}

\subsection{Problem Decomposition}

The heterogeneous HPC workflow scheduling problem is decomposed into two coupled subproblems:
(i) a \emph{task-node mapping problem}, which assigns each task to exactly one compute node under resource and feature constraints, and
(ii) a \emph{scheduling problem}, which determines task start and completion times subject to precedence and communication constraints.

In this work, quantum-inspired QUBO formulations are applied exclusively to the task-node mapping problem~\cite{sharma2025workflow}.
Scheduling metrics such as makespan and communication-aware execution time are computed using a classical schedule decoder applied to the resulting mappings.

\subsection{Problem Formulation}

Let $\mathcal{T} = \{t_1, \ldots, t_n\}$ denote tasks and $\mathcal{N} = \{n_1, \ldots, n_m\}$ denote compute nodes. Task $t_i$ requires $r_i$ CPU cores, feature set $F_i$, and executes in time $e_{ij}$ on node $n_j$ with capacity $c_j$ cores and features $G_j$. Precedence constraints form DAG with edges $(t_i, t_k)$ indicating $t_k$ depends on $t_i$. Communication time $d_{ik}$ applies when dependent tasks execute on different nodes.

Binary decision variables $x_{ij} \in \{0,1\}$ indicate whether task $t_i$ executes on node $n_j$. Continuous variables $s_i \geq 0$ denote task start times. Makespan $M$ equals maximum task completion time.

\textbf{Mapping Objective: Execution Cost Minimization}
\begin{equation}
\min \sum_{i=1}^{n} \sum_{j=1}^{m} e_{ij} x_{ij}
\end{equation}

\textbf{Constraint 1: Assignment Uniqueness}
\begin{equation}
\sum_{j=1}^m x_{ij} = 1 \quad \forall i \in \{1,\ldots,n\}
\end{equation}

\textbf{Constraint 2: Capacity Limits}
Node $n_j$ core usage never exceeds $c_j$ considering temporal task overlaps.

\textbf{Constraint 3: Feature Compatibility}
\begin{equation}
x_{ij} = 0 \quad \text{if } F_i \not\subseteq G_j
\end{equation}

\textbf{Constraint 4: Dependency Precedence}
\begin{equation}
s_k \geq s_i + e_{ij} x_{ij} + d_{ik}(1 - \delta_{jj'}) \quad \forall (t_i,t_k)
\end{equation}
where $\delta_{jj'}$ equals 1 if tasks assigned to same node.

\textbf{Constraint 5: Communication Overhead}
Communication time $d_{ik}$ included when $x_{ij} x_{kj'} = 1$ with $j \neq j'$.

If two dependent tasks are assigned to different nodes, a data transfer delay proportional to task data size and inter-node bandwidth is included.

\subsection{QUBO Formulation for Feature-Aware Task-Node Mapping}

We encode the problem through binary variables $x_{ij}$ and quadratic matrix $Q$ where energy $E(\mathbf{x}) = \mathbf{x}^T Q \mathbf{x}$. Variable indexing maps $(i,j)$ to position $v = i \cdot m + j$ in vector $\mathbf{x}$.

\textbf{Objective Term:}

\begin{equation}
E_{obj} = \sum_{i=1}^n \sum_{j=1}^m e_{ij} x_{ij}
\end{equation}

\textbf{Assignment Penalty:}
\begin{equation}
P_{assign} = \lambda_a \sum_{i=1}^n \left(1 - \sum_{j=1}^m x_{ij}\right)^2
\end{equation}

\scriptsize Expanding: $P_{assign} = \lambda_a \sum_{i=1}^n \left(1 - 2\sum_j x_{ij} + \sum_{j,j'} x_{ij}x_{ij'}\right)$
\\
\normalsize

\textbf{Capacity Penalty:}
\begin{equation}
P_{capacity} = \lambda_c \sum_{j=1}^m \left(\frac{\sum_{i=1}^n r_i x_{ij}}{c_j}\right)^2
\end{equation}

\textbf{Compatibility Penalty:}
\begin{equation}
P_{compat} = \lambda_{comp} \sum_{i=1}^n \sum_{j: F_i \not\subseteq G_j} x_{ij}
\end{equation}

\textbf{Dependency Penalty:}
\begin{equation}
P_{dep} = \lambda_d \sum_{(t_i,t_k)} \sum_j e_{ij} x_{ij}
\end{equation}

\textbf{Communication Penalty:}
\begin{equation}
P_{comm} = \lambda_{comm} \sum_{(t_i,t_k)} \sum_{j \neq j'} d_{ik} x_{ij} x_{kj'}
\end{equation}

\textbf{Total QUBO Energy:}
\begin{equation}
E_{\text{total}} = E_{\text{obj}} + P_{\text{assign}} + P_{\text{capacity}} + P_{\text{compat}} + P_{dep} + P_{comm}
\end{equation}

The QUBO formulation optimizes only the constrained task-node mapping.
Temporal scheduling aspects, including task precedence, communication delays, and makespan, are evaluated separately using a classical scheduling procedure.
This design choice avoids introducing time-indexed binary variables, which would significantly increase QUBO problem size and complexity.

\textbf{Penalty Coefficient Estimation:}
Let $e_{max} = \max_{i,j} e_{ij}$. We estimate:
\begin{align}
\lambda_a &= 3 \cdot e_{max} \cdot \alpha \\
\lambda_c &= 3 \cdot e_{max} \cdot \alpha \\
\lambda_{comp} &= 100 \cdot e_{max} \\
\lambda_{comm} &= 1 \cdot e_{max} \cdot \alpha
\end{align}
where $\alpha$ is penalty multiplier (default $\alpha = 1.0$).

\subsection{Algorithm Implementations}
\Cref{alg:milp} to \Cref{alg:qubo_qaoa} shown below:
\begin{algorithm}[t]
\caption{MILP Workflow Scheduling}
\label{alg:milp}
\begin{algorithmic}[1]
\STATE Create MILP model
\STATE Define binary $x_{ij}$ for compatible $(t_i,n_j)$
\STATE Define continuous $s_i$, makespan $M$
\STATE Minimize $M$
\STATE Add assignment: $\sum_j x_{ij} = 1$ for each $t_i$
\STATE Add capacity: $\sum_i r_i x_{ij} \leq c_j$ for each $n_j$
\STATE Add precedence: $s_k \geq s_i + e_{ij} x_{ij}$ for $(t_i,t_k)$
\STATE Add makespan: $M \geq s_i + e_{ij} x_{ij}$
\STATE Solve with branch and bound
\RETURN assignment, makespan, schedule
\end{algorithmic}
\end{algorithm}

\begin{algorithm}
\caption{CP-SAT Workflow Scheduling}
\label{alg:cpsat}
\begin{algorithmic}[1]
\STATE Create CP model
\STATE Define Boolean variables $x_{ij}$ for compatible $(t_i,n_j)$ pairs
\STATE Define IntVar $start_i$, $end_i$ for each task $t_i$
\STATE Define optional interval variables for each $(t_i,n_j)$
\STATE Add assignment constraint: $\sum_j x_{ij} = 1$ for each $t_i$
\STATE Add cumulative capacity constraints for each node
\STATE Add precedence constraints: $start_k \geq end_i$ for $(t_i,t_k)$
\STATE Define makespan variable $M = \max_i end_i$
\STATE Minimize $M$
\STATE Solve with CP-SAT solver
\STATE Extract assignment from solution
\RETURN makespan, utilization, schedule
\end{algorithmic}
\end{algorithm}

\begin{algorithm}[t]
\caption{Genetic Algorithm}
\label{alg:ga}
\begin{algorithmic}[1]
\STATE Initialize population of random task-node assignments
\FOR{generation $= 1$ to $G$}
    \STATE Evaluate fitness (negative makespan if feasible)
    \STATE Select parents via tournament selection
    \STATE Create offspring via crossover
    \STATE Apply mutation to offspring
    \STATE Replace population with offspring
\ENDFOR
\RETURN best individual from final population
\end{algorithmic}
\end{algorithm}

\begin{algorithm}
\caption{HEFT Workflow Scheduling}
\label{alg:heft}
\begin{algorithmic}[1]
\STATE Compute upward rank $rank_i$ for each task recursively
\STATE Sort tasks by decreasing rank
\FOR{each task $t_i$ in sorted order}
    \STATE $best\_node \gets null$, $best\_finish \gets \infty$
    \FOR{each node $n_j$ compatible with $t_i$}
        \STATE Compute ready time from dependencies
        \STATE Find earliest slot considering capacity
        \STATE Calculate finish time
        \IF{finish time $<$ $best\_finish$}
            \STATE Update $best\_node$, $best\_finish$
        \ENDIF
    \ENDFOR
    \STATE Assign $t_i$ to $best\_node$
\ENDFOR
\RETURN makespan, utilization, schedule
\end{algorithmic}
\end{algorithm}

\begin{algorithm}
\caption{QUBO Simulated Annealing}
\label{alg:qubo_sa}
\begin{algorithmic}[1]
\STATE Construct QUBO matrix $Q$ from objectives and penalties
\STATE Initialize binary vector $\mathbf{x}$ randomly
\STATE Set temperature $T \gets T_{init}$
\STATE $E_{best} \gets E(\mathbf{x})$, $\mathbf{x}_{best} \gets \mathbf{x}$
\FOR{iteration $= 1$ to $max\_iter$}
    \STATE Select random index $v$
    \STATE $\mathbf{x}' \gets \mathbf{x}$ with $x'_v = 1 - x_v$
    \STATE $E' \gets E(\mathbf{x}')$
    \IF{$E' < E$ OR $rand() < \exp((E - E') / T)$}
        \STATE $\mathbf{x} \gets \mathbf{x}'$, $E \gets E'$
        \IF{$E < E_{best}$}
            \STATE $\mathbf{x}_{best} \gets \mathbf{x}$, $E_{best} \gets E$
        \ENDIF
    \ENDIF
    \STATE $T \gets 0.95 \cdot T$
\ENDFOR
\STATE Decode $\mathbf{x}_{best}$ to assignment
\STATE Decode mapping to schedule using classical scheduler
\RETURN assignment, makespan, utilization
\end{algorithmic}
\end{algorithm}
\begin{algorithm}[t]
\caption{QUBO Quantum Annealing (Multi-Attempt)}
\label{alg:qubo_qa}
\begin{algorithmic}[1]
\STATE Initialize best solution, best energy
\FOR{read $= 1$ to $N_{reads}$}
    \STATE Randomize $T_{init} \in [80, 120]$
    \STATE Randomize $\alpha \in [0.5, 1.5]$
    \STATE Run QUBO-SA with randomized parameters
    \IF{energy < best energy}
        \STATE Update best solution
    \ENDIF
\ENDFOR
\RETURN best solution from all reads
\end{algorithmic}
\end{algorithm}

\begin{algorithm}[t]
\caption{QUBO QAOA (Layered Optimization)}
\label{alg:qubo_qaoa}
\begin{algorithmic}[1]
\STATE Initialize penalty multipliers for $p$ layers
\STATE $[\alpha_1, \alpha_2, \ldots, \alpha_p] = [0.4, 0.7, 1.0, 1.3]$
\FOR{layer $= 1$ to $p$}
    \STATE Set $T_{init} = 60 + \alpha_{layer} \cdot 30$
    \STATE Run QUBO-SA with layer penalty
    \STATE Store result
\ENDFOR
\STATE Select best feasible result across layers
\RETURN best solution
\end{algorithmic}
\end{algorithm}

\subsection{Enhancement Strategies}

\textbf{Strategy 1: Adaptive Penalty Estimation}

Solve instance with CP-SAT to obtain optimal makespan $M^*$. Estimate penalties as $\lambda = 2.5 \cdot M^*$ rather than $3 \cdot e_{max}$. Run QUBO-SA with estimated penalties. This approach uses exact solver analysis to guide penalty calibration.

\textbf{Strategy 2: Hybrid QUBO with Repair}

Attempt QUBO-SA with weak penalties ($\alpha = 0.5$) for fast optimization. If solution infeasible, repair using HEFT. This combines optimization quality of QUBO with guaranteed feasibility of classical heuristics. Hybrid approach trades slight complexity for reliability.

\textbf{Strategy 3: Multi Attempt Annealing}

Execute 20 QUBO-SA runs with randomly varied initial temperatures (80 to 120) and penalty multipliers (1.5 to 3.0). Select solution with lowest energy among feasible results. This automated parameter exploration mimics quantum annealing behavior through classical sampling.

\textbf{Strategy 4: Progressive Penalty Strengthening}

Start with weak penalty multiplier $\alpha = 1.0$. If solution infeasible, increase $\alpha \gets 1.5 \cdot \alpha$ and retry. Repeat up to 5 iterations until feasible solution found. This iterative approach systematically explores penalty coefficient space.

\textbf{Strategy 5: Two Stage Optimization}

Stage 1 employs QUBO-SA with strong penalties ($\alpha = 3.0$) focusing on constraint satisfaction. Stage 2 uses HEFT as fallback if QUBO fails. This separates constraint enforcement from objective optimization.

\subsection{Experimental Design}

We conduct four experiments designed to expose feasibility, robustness, and scaling behavior under increasing constraint complexity.

\textbf{Experiment 0: Progressive Constraint Addition (Medium\_6T\footnote{\scriptsize Medium\_6T - a medium-complexity synthetic workflow (6 tasks) that simultaneously activates all five hard constraints and serves as the primary instance for progressive constraint and penalty sensitivity analysis.}).}
We activate constraints incrementally to identify when feasibility breaks for each solver:
(i) objective only, (ii) +assignment uniqueness, (iii) +capacity, (iv) +feature compatibility, (v) +dependencies, (vi) +communication.
At each step we record feasibility, makespan, utilization, and runtime for all seven algorithms (MILP, CP-SAT, GA, HEFT, QUBO-SA, QUBO-MultiSA, QUBO-MultiSAOA).

\textbf{Experiment 1: Penalty Sensitivity (Medium\_6T).}
We evaluate penalty calibration difficulty by sweeping the QUBO-SA penalty multiplier over
$\{0.05, 0.1, 0.5, 1.0, 2.0, 5.0\}$ with three runs per setting.
We report feasibility rate and, for feasible outputs, makespan and utilization.
Classical solvers (MILP, CP-SAT, GA, HEFT) and the other QUBO variants are included as reference points.

\textbf{Experiment 2: Ground-Truth Validation (W1/W2).}
We validate correctness using two small workflows with known optimal makespan of 10.0\,s:
W1 (3-task chain) and W2 (4-task diamond DAG).
All seven algorithms are evaluated to ensure they recover feasibility and the expected makespan~\cite{sharma2025workflow}.

\textbf{Experiment 3: Scaling (Synthetic 5-20 tasks).}
We evaluate feasibility and performance at $|T|\in\{5,10,15,20\}$ with three random instances per scale.
All seven algorithms are tested under the same limits and evaluation pipeline.
We report feasibility rate, makespan (for feasible outputs), utilization, and runtime.

\subsection{System Configuration}

Experiments executed on Ubuntu 24.04 system with Python 3.11. Classical solvers: PuLP 2.8 with CBC, OR-Tools CP-SAT 9.11. QUBO-SA parameters: 10000 iterations, initial temperature 100, cooling rate 0.95. GA parameters: population 100, generations 500. Timeout limit 300 seconds for exact solvers. For deterministic configurations (QUBO-SA, CP-SAT, MILP, HEFT), random seeds are fixed for reproducibility.
For QUBO-MultiSA and QUBO-MultiSAOA, seeds are intentionally varied across runs as part of controlled stochastic sampling; the base seed is fixed so the entire ensemble is reproducible.
All QUBO-based methods were implemented using a custom Python research codebase developed for this study.
Simulated annealing was implemented as a classical baseline for QUBO optimization using a standard Metropolis update rule and geometric cooling schedule, with identical formulations shared across all QUBO variants.
We emphasize that differences between QUBO-SA, QUBO-MultiSA, and QUBO-MultiSAOA reflect solution strategies rather than distinct formulations; all methods operate on the same penalty-based QUBO model.

\section{Experimental Results}

This section reports experimental results for four complementary experiments designed to evaluate correctness, feasibility behavior, penalty sensitivity, and scalability of quantum-inspired QUBO methods relative to classical baselines.
Unless stated otherwise, all reported makespan and utilization values are derived from a unified classical schedule decoder applied to solver-produced task-node mappings.
Feasibility is reported explicitly; objective values are reported only for feasible solutions.

\paragraph{QUBO Solver Variants.}
For clarity, we define three quantum-inspired QUBO solver variants evaluated throughout the experiments, all operating on the same underlying penalty-based QUBO formulation but differing in solution strategy.
\textit{QUBO-SA} denotes a baseline simulated annealing solver applied to a fixed penalty-based QUBO formulation.
\textit{QUBO-MultiSA} extends this baseline by performing multiple independent annealing runs with varied random seeds and selecting the best feasible solution.
\textit{QUBO-MultiSAOA} further augments this approach using structured parameter variation inspired by alternating-operator schedules, aiming to improve robustness under interacting constraints.
Unless stated otherwise, all QUBO variants operate on the same underlying formulation and differ only in solution strategy.

\subsection{Experiment 0: Progressive Constraint Addition}

To study feasibility degradation as constraint structure increases, we progressively activate constraints on a Medium\_6T instance.
Starting from an objective-only formulation, constraints are added in the following order:
assignment uniqueness, capacity limits, feature compatibility, task dependencies, and finally communication costs.

\begin{figure}[t]
    \centering
    \includegraphics[width=\linewidth]{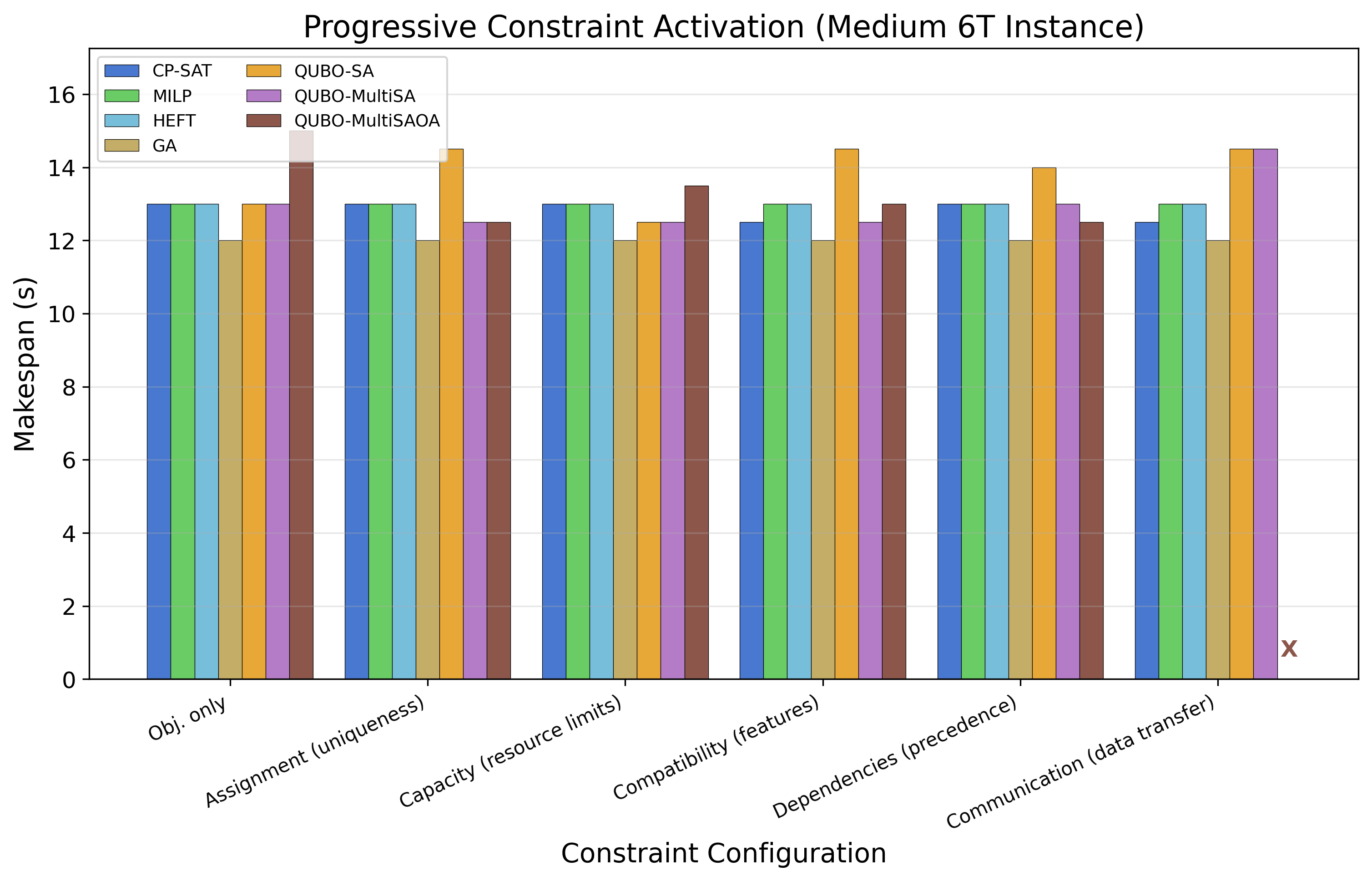}
    \caption{Makespan per solver under progressive constraint activation (Experiment~0, Medium\_6T instance).
    Constraints are added incrementally from objective-only to full communication-aware scheduling.
    Red-brown X marks infeasibility: QUBO-MultiSAOA fails when communication constraints are introduced.
    At full constraints, CP-SAT achieves 12.5\,s, GA 12.0\,s, MILP/HEFT 13.0\,s, and QUBO-SA/MultiSA 14.5\,s.}
    \label{fig:progressive}
\end{figure}

All solvers remain feasible through the dependency stage.
When communication constraints are activated, QUBO-MultiSAOA becomes infeasible on this instance, while all remaining solvers retain feasibility.
This indicates that communication-aware scheduling constitutes a critical stressor for certain QUBO variants as illustrated in \Cref{fig:progressive}.

For the fully constrained instance, CP-SAT achieves a makespan of 12.5\,s, GA 12.0\,s, MILP and HEFT 13.0\,s, and QUBO-SA/QUBO-MultiSA 14.5\,s.
QUBO-MultiSAOA produces no feasible solution and is reported as infeasible.
These results demonstrate that QUBO feasibility can degrade as interacting constraints accumulate, even when classical solvers remain robust.

\subsection{Experiment 1: Penalty Sensitivity Analysis}

\subsubsection{Test Trail 1: Penalty Coefficient Sensitivity}

\Cref{tab:penalty} presents penalty sensitivity analysis demonstrating quantitative calibration requirements. Classical methods (MILP, CP-SAT, GA, HEFT) achieve perfect feasibility independent of penalty considerations. QUBO-SA exhibits strong sensitivity to penalty multiplier selection.

\begin{table}[t]
\caption{Penalty Coefficient Sensitivity on 6 Task Instance}
\label{tab:penalty}
\centering
\small
\begin{tabular}{@{}lccc@{}}
\toprule
Method & Feasible & Makespan (s) & Runtime (ms) \\
\midrule
\multicolumn{4}{l}{\textit{Classical Baselines}} \\
MILP & 1/1 & 13.0 & 41.2 \\
CP-SAT & 1/1 & 13.0 & 29.6 \\
GA & 1/1 & 12.0 & 12912.2 \\
HEFT & 1/1 & 13.0 & 2.0 \\
\midrule
\multicolumn{4}{l}{\textit{QUBO Penalty Sensitivity}} \\
$\alpha = 0.05$ & 0/3 & N/A & N/A \\
$\alpha = 0.10$ & 0/3 & N/A & N/A \\
$\alpha = 0.50$ & 1/3 & 13.0 & 26.3 \\
$\alpha = 1.00$ & 3/3 & 13.3 & 25.8 \\
$\alpha = 2.00$ & 3/3 & 14.0 & 26.6 \\
$\alpha = 5.00$ & 3/3 & 13.2 & 26.6 \\
\midrule
\multicolumn{4}{l}{\textit{QUBO Variants}} \\
QUBO-MultiSA & 3/3 & 12.5 & 1058.5 \\
QUBO-MultiSAOA & 3/3 & 13.3 & 48.2 \\
\bottomrule
\end{tabular}
\end{table}

QUBO-SA exhibits a sharp feasibility transition with $\alpha$: $0/3$ feasible runs at $\alpha \in \{0.05, 0.10\}$, $1/3$ at $\alpha=0.5$, and $3/3$ for $\alpha \ge 1.0$, defining a clear effective calibration regime. QUBO-MultiSA and QUBO-MultiSAOA both attain $100\%$ feasibility through automated multi-attempt and layered parameter exploration (average makespan $12.5$\,s and $13.3$\,s respectively), showing that systematic parameter search substantially reduces reliance on manual penalty tuning.

\begin{figure}[t]
    \centering
    \includegraphics[width=\linewidth]{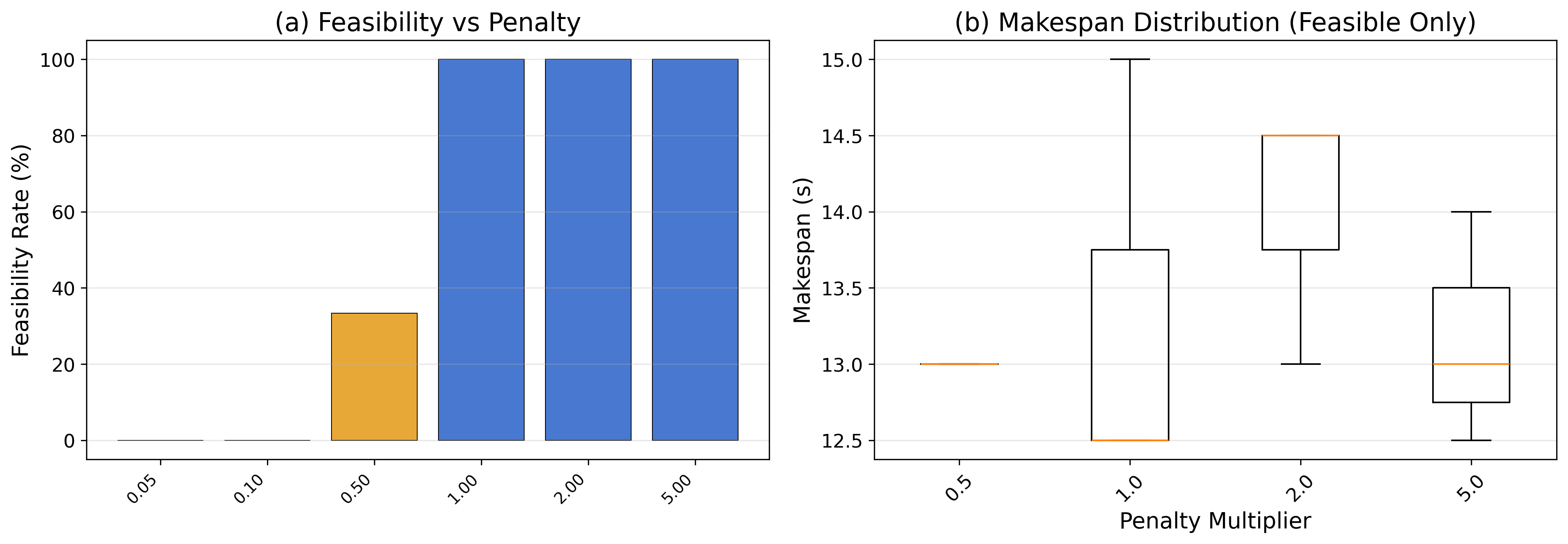}
    \caption{Boxplots of QUBO-SA penalty sensitivity. \textbf{Left:} feasibility rate by penalty multiplier. \textbf{Right:} makespan distribution across 3 runs per feasible penalty setting, showing moderate stochastic variation.}
    \label{fig:penalty_box}
\end{figure}

\Cref{fig:penalty_box} presents boxplots of makespan variation across the three runs per penalty setting, illustrating that stochastic variation is moderate once feasibility is achieved. For feasible runs QUBO-SA runtime remains stable at $\sim$26\,ms with memory usage below $0.02$\,MB.

\subsubsection{Test Trail 2: Sequential Constraints}
\begin{figure}[t]
    \centering
    \includegraphics[width=\linewidth]{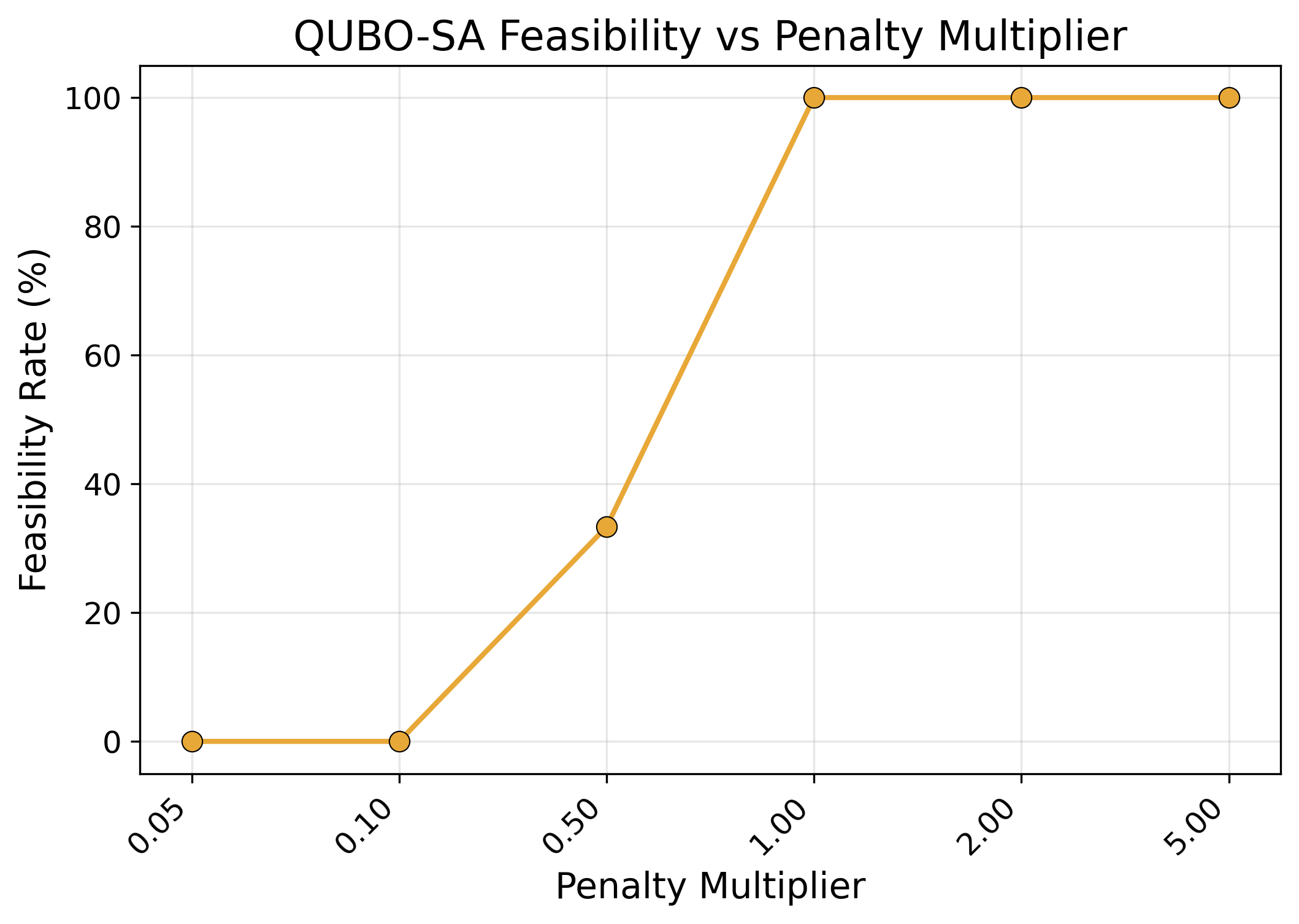}
    \vspace{1mm}
    \includegraphics[width=\linewidth]{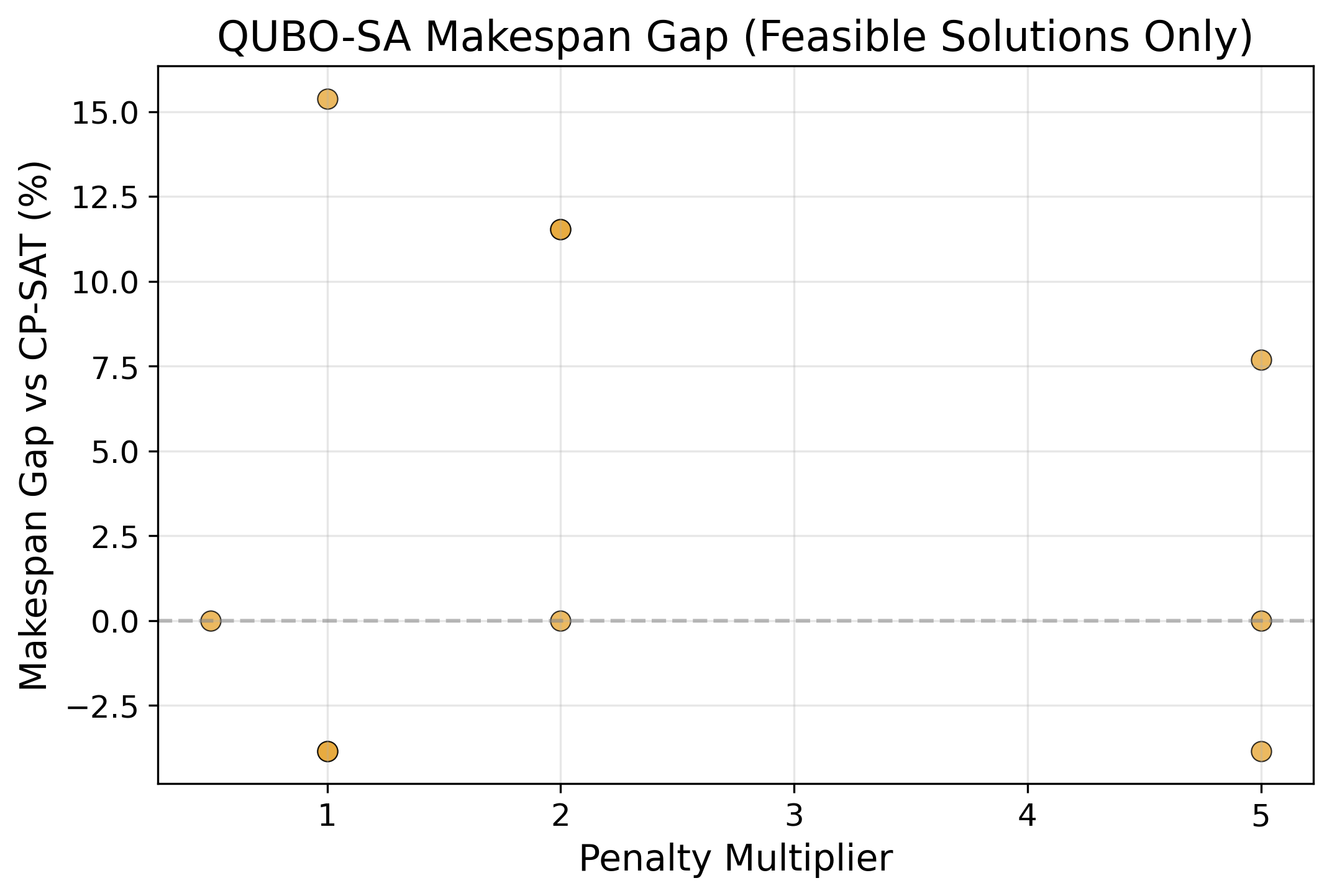}
    \caption{Penalty sensitivity of QUBO-SA on the Medium\_6T instance (Experiment~1).
    \textbf{Top:} feasibility rate vs penalty multiplier.
    \textbf{Bottom:} makespan gap relative to the best feasible result.}
    \label{fig:penalty}
\end{figure}

\Cref{fig:penalty} visualizes the same penalty sweep, with the feasibility curve (top) and the makespan gap relative to the best feasible result (bottom), making the sharp transition between weak and moderate penalties explicit.

These results confirm that penalty calibration is not optional: insufficient penalties lead to systematic infeasibility, while overly strong penalties distort optimization without improving feasibility~\cite{hadfield2019quantum}.

\subsection{Experiment 2: Ground-Truth Validation (W1, W2)}
\subsubsection{Validation Test 1 - Standard Sample Test~\cite{sharma2025workflow}}
\Cref{tab:val-energy} presents the validation benchmark results for the three baseline implementations (CP-SAT, HEFT, QUBO-SA), including QUBO-SA energy values that confirm proper penalty calibration. CP-SAT achieves perfect correctness on both W1 and W2 with the exact 10.0\,s makespan in $\sim$12\,ms, verified optimal through exhaustive state space examination. HEFT likewise reaches the same 10.0\,s makespan in under $0.5$\,ms, demonstrating the efficiency of priority-based greedy heuristics on these carefully constructed validation benchmarks.

QUBO-SA achieves $100\%$ feasibility across all 5 runs per instance and recovers the exact 10.0\,s optimum on every run, with an average runtime of $2.8$\,ms (slightly slower than HEFT but substantially faster than CP-SAT). Energy values ranging from $-32.4$ to $-48.0$ indicate constraint satisfaction with penalties dominating the objective, confirming proper penalty calibration.

\begin{table}[t]
\caption{Baseline validation with QUBO-SA energy values (Expected Makespan: 10.0\,s).}
\label{tab:val-energy}
\centering
\small
\begin{tabular}{@{}lcccc@{}}
\toprule
Method & Feasible & Makespan (s) & Runtime (ms) & Energy \\
\midrule
\multicolumn{5}{l}{\textit{W1 Linear (3 tasks)}} \\
CP-SAT & 1/1 & 10.0 & 15.9 & N/A \\
HEFT & 1/1 & 10.0 & 0.44 & N/A \\
QUBO-SA & 5/5 & 10.0 & 2.7 & -34.0 \\
\midrule
\multicolumn{5}{l}{\textit{W2 Diamond (4 tasks)}} \\
CP-SAT & 1/1 & 10.0 & 8.4 & N/A \\
HEFT & 1/1 & 10.0 & 0.39 & N/A \\
QUBO-SA & 5/5 & 10.0 & 2.9 & -46.9 \\
\bottomrule
\end{tabular}
\end{table}

The systematic success of CP-SAT (verified optimal), HEFT (greedy upward rank), and QUBO-SA (penalty multiplier $\alpha=1.0$) across both linear and branching workflow structures confirms proper encoding of all five hard constraints.

\subsubsection{Validation Test 2 - Solution Strategy Comparison}

\Cref{tab:solutions} compares the baseline methods and the five enhancement strategies on the W2 instance. All approaches reach the optimal $10.0$\,s makespan with $100\%$ feasibility, reflecting that the 4-task / 3-node workflow leaves enough solution space for multiple optimization approaches to converge.

\subsubsection{Validation Test 3 - Enhancement Strategies}
Among the enhancement strategies in \Cref{tab:solutions}, all five (adaptive penalty estimation guided by CP-SAT, hybrid QUBO/HEFT repair, multi-attempt annealing, progressive penalty strengthening, and two-stage optimization) maintain perfect feasibility on W2 with runtimes between $2.5$ and $3.7$\,ms.

\begin{table}[t]
\caption{Solution Strategy Comparison on W2 Instance}
\label{tab:solutions}
\centering
\small
\begin{tabular}{@{}lccc@{}}
\toprule
Strategy & Feasible & Makespan (s) & Runtime (ms) \\
\midrule
\multicolumn{4}{l}{\textit{Baselines}} \\
CP-SAT & 1/1 & 10.0 & 8.3 \\
HEFT & 1/1 & 10.0 & 0.4 \\
QUBO-SA & 5/5 & 10.0 & 3.3 \\
\midrule
\multicolumn{4}{l}{\textit{Enhancement Strategies}} \\
Adaptive Penalty & 5/5 & 10.0 & 3.7 \\
Hybrid Repair & 5/5 & 10.0 & 2.9 \\
Multi Attempt & 5/5 & 10.0 & 3.0 \\
Progressive Penalty & 5/5 & 10.0 & 2.8 \\
Two Stage & 5/5 & 10.0 & 2.5 \\
\bottomrule
\end{tabular}
\end{table}

The uniform success across all methods on this 4 task instance demonstrates that properly formulated QUBO approaches achieve competitive performance with classical methods. Enhancement strategies maintain reliability without significant runtime overhead. Hybrid repair and multi attempt approaches provide automatic parameter adaptation that could prove valuable on more challenging instances.

\subsubsection{Overall Validation Tests Evaluation}
We first validate correctness using two small workflows with analytically known optimal solutions:
W1 (3-task linear chain) and W2 (4-task diamond DAG).
Both instances have an expected optimal makespan of 10.0\,s.
This experiment serves as a sanity check: failure to recover the optimal makespan indicates a formulation or implementation error rather than an algorithmic limitation.

Table~\ref{tab:validation} shows that all seven algorithms, MILP, CP-SAT, GA, HEFT, QUBO-SA, QUBO-MultiSA, and QUBO-MultiSAOA, produce feasible schedules and recover the expected 10.0\,s makespan on both workflows.
This confirms that all formulations and decoders are implemented correctly and that QUBO-based approaches are capable of solving small, well-structured instances.

\begin{table}[t]
\caption{Ground-truth validation on W1 and W2 (expected makespan: 10.0\,s).}
\label{tab:validation}
\centering
\scriptsize
\begin{tabular}{@{}llccc@{}}
\toprule
Workflow & Solver & Feasible & Makespan (s) & Runtime (ms) \\
\midrule
\multirow{7}{*}{W1 (3 tasks)}
& MILP & 1/1 & 10.0 & 27.5 \\
& CP-SAT & 1/1 & 10.0 & 8.3 \\
& GA & 3/3 & 10.0 & 9325.5 \\
& HEFT & 3/3 & 10.0 & 1.5 \\
& QUBO-SA & 5/5 & 10.0 & 24.6 \\
& QUBO-MultiSA & 3/3 & 10.0 & 1004.7 \\
& QUBO-MultiSAOA & 2/2 & 10.0 & 46.2 \\
\midrule
\multirow{7}{*}{W2 (4 tasks)}
& MILP & 1/1 & 10.0 & 31.4 \\
& CP-SAT & 1/1 & 10.0 & 10.9 \\
& GA & 3/3 & 10.0 & 10878.9 \\
& HEFT & 3/3 & 10.0 & 1.6 \\
& QUBO-SA & 5/5 & 10.0 & 25.7 \\
& QUBO-MultiSA & 3/3 & 10.0 & 1096.6 \\
& QUBO-MultiSAOA & 2/2 & 10.0 & 47.7 \\
\bottomrule
\end{tabular}
\end{table}

\subsection{Experiment 3: Scaling Behavior (5-20 Tasks)}

Experiment~3 evaluates solver robustness and performance as problem size increases.
We consider synthetic workflow instances with 5, 10, 15, and 20 tasks, using three random instances per scale and identical solver limits.

\subsubsection{Scaling Test 1 - Performance Trends}

Table~\ref{tab:scaling} reports aggregate performance metrics on feasibility, average makespan, and runtime i.e., for selected representative methods across all scales.
CP-SAT remains feasible on all instances and provides near-optimal schedules, with average makespan increasing from 9.2\,s at 5 tasks to 17.4\,s at 20 tasks.
Runtime grows modestly from 15\,ms to 41\,ms, remaining practical despite worst-case exponential complexity.

HEFT also achieves perfect feasibility at all scales, with makespan closely tracking CP-SAT and consistently minimal runtimes below 1.5\,ms.
This confirms the strong scalability of greedy heuristics for heterogeneous workflow scheduling.

The hybrid QUBO-classical repair strategy maintains perfect feasibility across all scales.
Its makespan is slightly higher than CP-SAT at small sizes but converges to HEFT-quality solutions at larger scales, while runtime remains below 3\,ms.
Multi-attempt annealing also preserves feasibility but exhibits higher makespan and increased runtime due to repeated QUBO-SA executions.

\begin{table}[t]
\caption{Scaling performance across 5-20 tasks (3 instances per scale).}
\label{tab:scaling}
\centering
\scriptsize
\begin{tabular}{@{}lcccc@{}}
\toprule
Tasks & Method & Feasible & Makespan (s) & Runtime (ms) \\
\midrule
\multirow{4}{*}{5}
& CP-SAT & 3/3 & 9.2 & 15.0 \\
& HEFT & 3/3 & 9.1 & 0.5 \\
& Hybrid & 9/9 & 10.4 & 3.0 \\
& Multi Attempt & 9/9 & 10.1 & 3.0 \\
\midrule
\multirow{4}{*}{10}
& CP-SAT & 3/3 & 10.6 & 34.0 \\
& HEFT & 3/3 & 9.8 & 1.0 \\
& Hybrid & 9/9 & 13.6 & 3.0 \\
& Multi Attempt & 9/9 & 13.0 & 3.0 \\
\midrule
\multirow{4}{*}{15}
& CP-SAT & 3/3 & 18.2 & 36.0 \\
& HEFT & 3/3 & 17.8 & 1.0 \\
& Hybrid & 9/9 & 17.8 & 1.0 \\
& Multi Attempt & 9/9 & 21.9 & 4.0 \\
\midrule
\multirow{4}{*}{20}
& CP-SAT & 3/3 & 17.4 & 41.0 \\
& HEFT & 3/3 & 16.5 & 1.0 \\
& Hybrid & 9/9 & 16.5 & 1.0 \\
& Multi Attempt & 9/9 & 20.9 & 5.0 \\
\bottomrule
\end{tabular}
\end{table}

\subsubsection{Scaling Test 2 - Feasibility Regimes}

\begin{figure}[t]
    \centering
    \includegraphics[width=\linewidth]{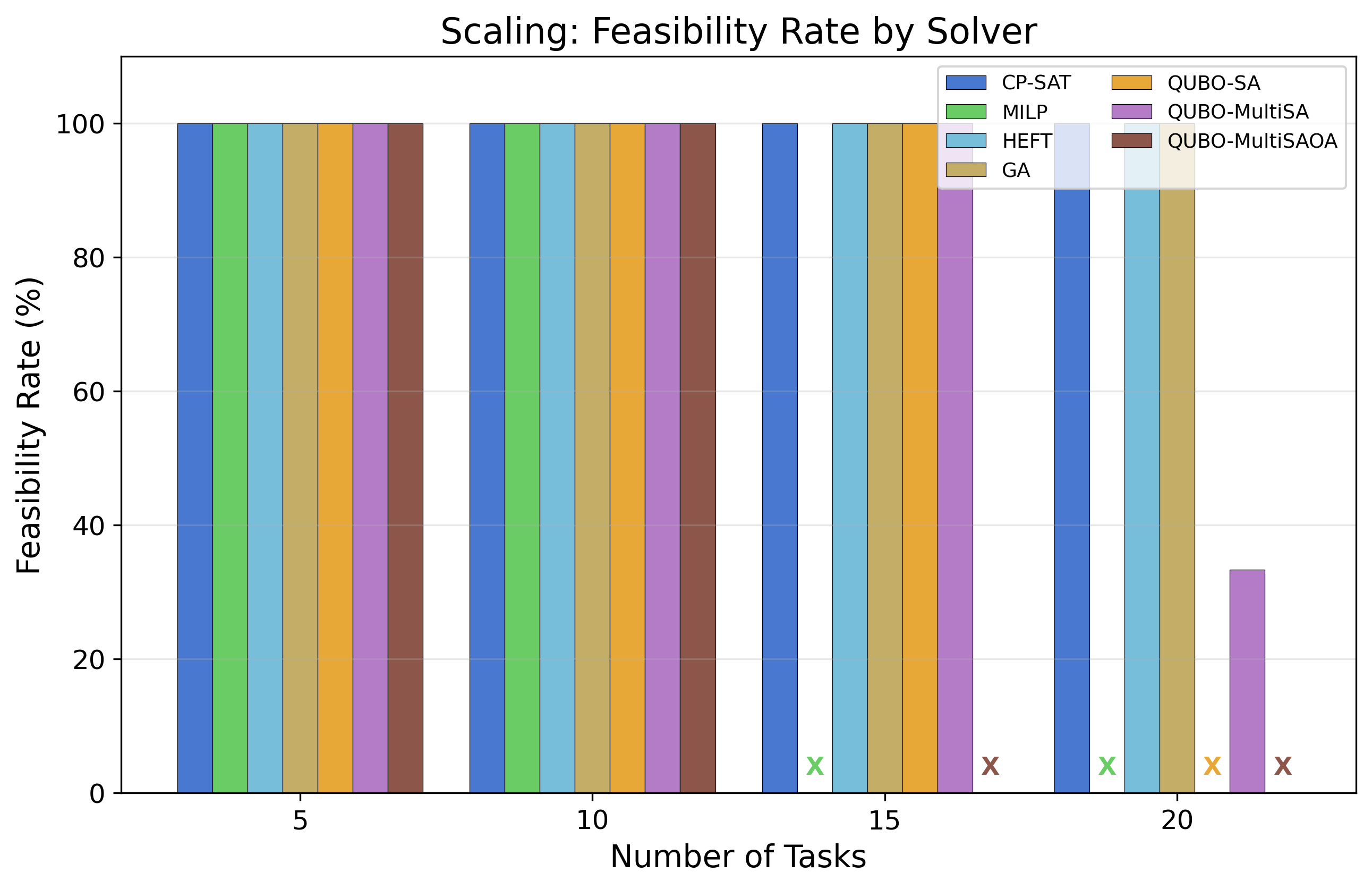}
    \vspace{1mm}
    \includegraphics[width=\linewidth]{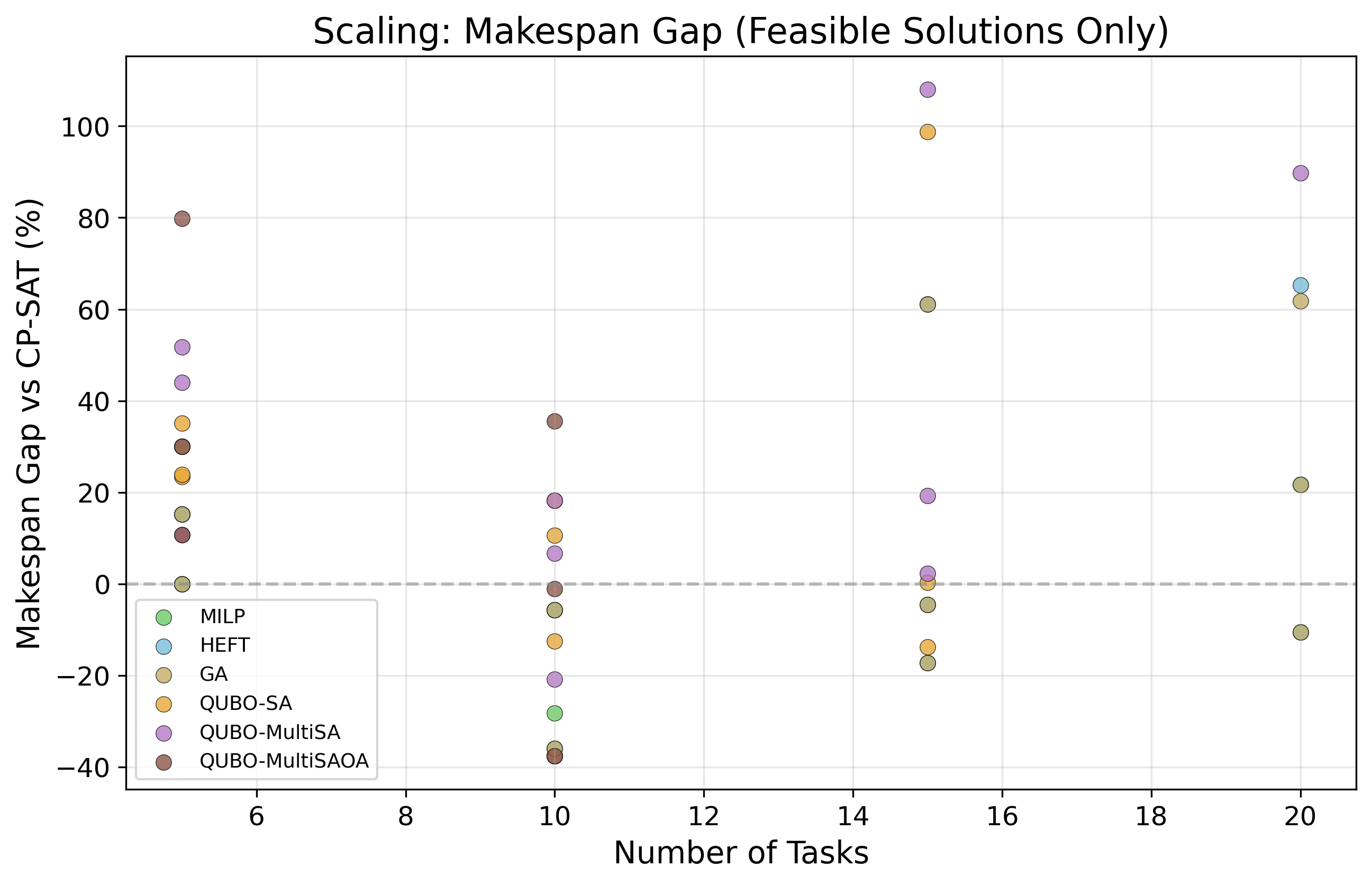}
    \caption{Scaling behavior across 5-20 tasks.
    \textbf{Top:} feasibility rate as a function of problem size.
    \textbf{Bottom:} makespan gap relative to CP-SAT for feasible solutions only.
    Classical solvers remain robust across all scales, while QUBO variants exhibit feasibility degradation as problem size increases.}
    \label{fig:scaling}
\end{figure}

Figure~\ref{fig:scaling} summarizes feasibility trends and solution quality gaps across all seven algorithms.
CP-SAT, GA, and HEFT remain feasible at all tested scales.
MILP is feasible up to 10 tasks but fails at 15 and 20 tasks, indicating a practical scalability boundary under the configured limits.

QUBO-based methods exhibit intermediate behavior.
QUBO-SA remains feasible through 15 tasks but fails completely at 20 tasks.
QUBO-MultiSA remains feasible through 15 tasks and achieves partial feasibility at 20 tasks.
QUBO-MultiSAOA becomes infeasible beyond 15 tasks.
Among feasible outputs, HEFT and CP-SAT achieve the lowest runtimes, while QUBO variants incur higher overhead and larger makespan gaps at increasing scale.

\begin{figure}[t]
    \centering
    \includegraphics[width=\linewidth]{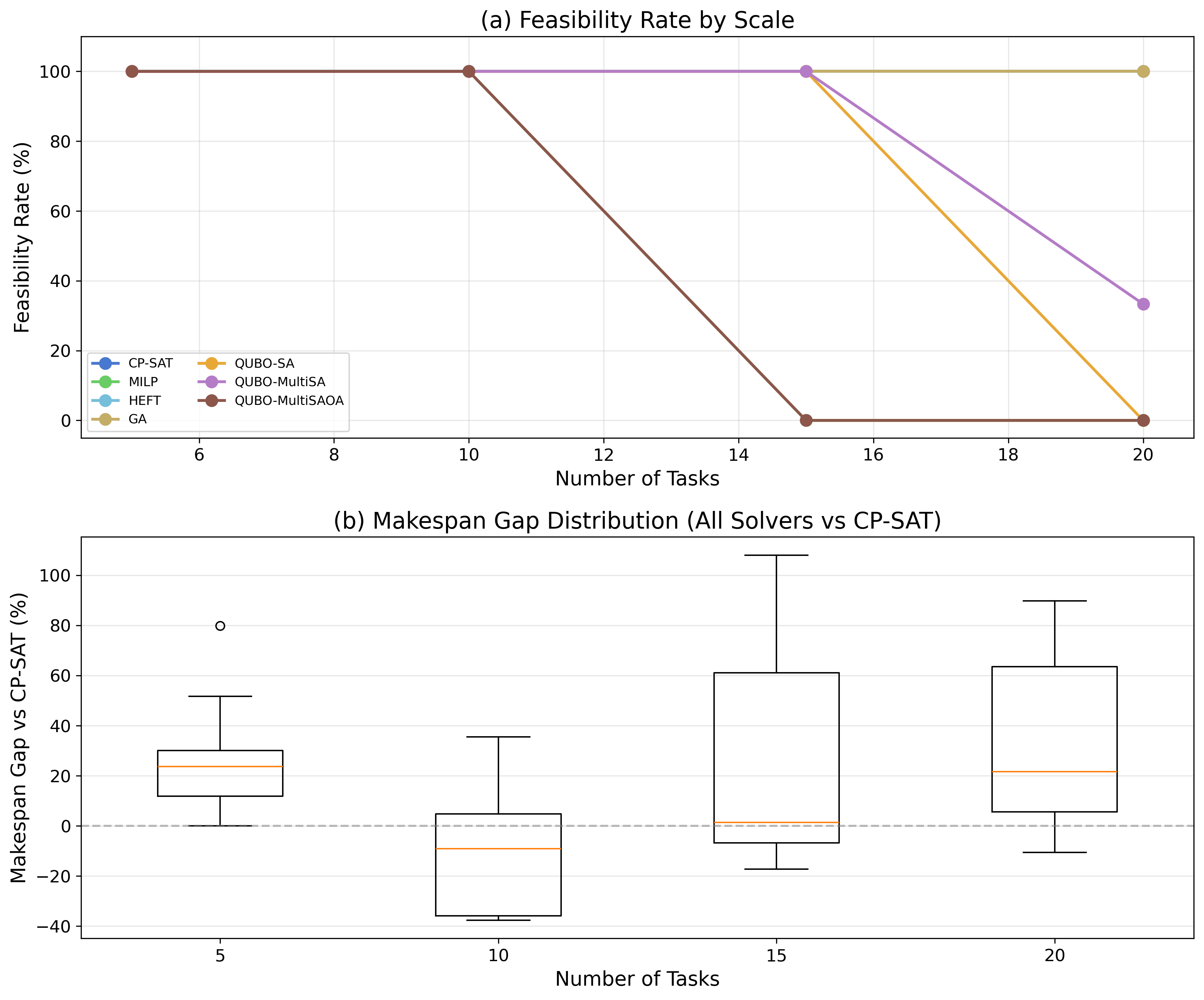}
    \caption{Scaling behavior across 5 to 20 tasks (3 instances per scale). \textbf{Top:} feasibility rate per solver, using consistent naming with Fig.~4 (QUBO-MultiSA = multi-attempt annealing, QUBO-EnsembleSA = layered QAOA-inspired). \textbf{Bottom:} per-instance makespan gap relative to CP-SAT for feasible solutions only, showing increasing variance for QUBO variants at larger scales.}
    \label{fig:scaling_box}
\end{figure}

\Cref{fig:scaling_box} provides boxplots of the makespan gap relative to CP-SAT across scales, confirming that QUBO methods exhibit both lower feasibility and higher solution quality variance at larger problem sizes.

Overall, these results show that quantum-inspired QUBO methods remain competitive only in small-to-moderate regimes, whereas classical solvers and heuristics dominate as constraint interactions and problem size increase.

\section{Discussion}

This section interprets the experimental findings with respect to the research questions and evaluates the practical viability of quantum-inspired QUBO approaches for heterogeneous HPC scheduling.

\subsection{Correctness and Validation}

Ground-truth validation on W1 and W2 confirms the correctness of all solver implementations.
All seven methods recover the expected optimal makespan of 10.0\,s, verifying correct handling of assignment, capacity, feature compatibility, precedence, and communication on small, structured workflows.
These results serve strictly as sanity checks: they establish formulation and implementation correctness but do not imply scalability or general performance advantages.

Crucially, validation confirms that the QUBO formulation correctly encodes \emph{mapping-level constraints}, while precedence and communication effects are consistently enforced during classical schedule derivation.
This decomposition is essential to maintain tractable QUBO sizes and reflects standard practice in hybrid quantum-classical optimization.

\subsection{Impact of Constraint Structure (RQ1)}

Progressive constraint activation shows that feasibility is strongly governed by constraint structure.
All solvers remain feasible through dependency constraints, indicating that precedence alone does not fundamentally challenge QUBO-based mapping.
In contrast, the introduction of communication constraints causes feasibility loss for specific QUBO variants, most notably QUBO-MultiSAOA on the Medium\_6T instance.

Communication costs introduce non-local interactions that are difficult to balance using quadratic penalties alone.
While classical solvers and heuristics remain robust, QUBO feasibility becomes sensitive to both penalty calibration and constraint interaction.
These results demonstrate that QUBO competitiveness is dictated by the \emph{constraint mix}, not merely by problem size, with communication-heavy workloads constituting a critical stress regime.

\subsection{Penalty Calibration and Trade-offs (RQ2)}

Penalty sensitivity analysis identifies penalty calibration as the primary determinant of QUBO feasibility.
Weak penalties systematically yield infeasible solutions, whereas sufficiently strong penalties restore feasibility at the cost of reduced objective sensitivity.
The observed sharp feasibility transition indicates threshold behavior rather than a smooth trade-off.

This exposes a fundamental limitation of penalty-based QUBO formulations: constraint enforcement introduces an additional calibration dimension absent in MILP and CP-SAT.
As a result, automated or adaptive penalty strategies are essential for practical use.
Among the evaluated approaches, hybrid repair is particularly effective, achieving high feasibility with minimal overhead and demonstrating that classical heuristics can complement QUBO optimization without negating its value.

\subsection{Scalability and Empirical Regimes (RQ3)}

Scaling experiments reveal distinct regimes of solver applicability.
CP-SAT and HEFT remain feasible across all tested scales up to 20 tasks, while MILP encounters a practical feasibility boundary between 10 and 15 tasks.
QUBO-based methods exhibit intermediate behavior: QUBO-SA remains feasible up to 15 tasks but fails at 20 tasks, QUBO-MultiSA retains partial feasibility at the largest scale, and QUBO-MultiSAOA becomes infeasible beyond 15 tasks.

\subsection{Empirical Evidence Toward Advantage (RQ4)}
Across the evaluated instances and solver budgets, we do not observe a consistent advantage of QUBO-based methods over CP-SAT (optimality/robustness) or HEFT (speed). QUBO approaches remain competitive in small-to-moderate regimes when penalties are well calibrated or augmented with repair, but the additional calibration effort and feasibility degradation at larger scales prevent an operational advantage under our current setup.

These results indicate that, under current formulations and solvers, QUBO approaches do not surpass classical methods in scalability or robustness.
However, they remain competitive in small-to-moderate regimes when supported by enhancement strategies.
The absence of a clear quantum advantage is consistent with broader findings in quantum optimization research and reinforces the need for rigorous empirical evaluation.

\subsection{Practical Implications for HPC Scheduling}

In practice, CP-SAT is best suited for small workflows requiring guaranteed optimality, while HEFT remains the most efficient choice for larger workflows where near-optimal solutions suffice.
Quantum-inspired QUBO methods combined with hybrid repair occupy an intermediate position, offering feasible solutions with modest overhead and serving as a structured testbed for hybrid quantum-classical mapping and scheduling pipelines.

At present, the primary value of QUBO formulations lies not in outperforming classical solvers but in providing a principled pathway toward hybrid quantum-HPC optimization.
As quantum hardware matures, such formulations may enable acceleration of the most combinatorially challenging subproblems within larger scheduling workflows.

\section{Conclusion}

This work presents a systematic and reproducible evaluation of quantum-inspired QUBO formulations for heterogeneous HPC workflow scheduling. By decomposing the problem into task-node mapping and classical schedule derivation, we ensure a mathematically consistent formulation and enable fair comparison with classical exact and heuristic solvers.

Ground-truth validation, progressive constraint activation, penalty sensitivity analysis, and scaling experiments jointly characterize the feasibility, robustness, and performance limits of QUBO-based approaches. Properly calibrated QUBO formulations are correct on small instances and remain competitive at moderate scales when supported by enhancement strategies, but classical methods (MILP and CP-SAT for optimality, HEFT for speed) remain dominant across the tested sizes.

Quantum-inspired QUBO methods therefore do not yet provide a clear performance advantage under current formulations and solver implementations. They occupy an intermediate regime in which feasibility and solution quality can be maintained at the cost of additional calibration effort, consistent with the broader state of quantum optimization research and underscoring the need for empirical rigor over speculative advantage claims.

The primary contribution of this work is a structured methodology for mapping realistic, multi-constraint scheduling problems into QUBO form, evaluating feasibility under increasing constraint complexity, and identifying empirical regimes of competitiveness. This framework establishes a sound foundation for future research on hybrid quantum-classical scheduling systems.

\subsection*{Limitations}
Several limitations apply.
\textit{(i)}~Problem sizes (5 to 20 tasks) are small relative to production HPC workflows, which may involve hundreds or thousands of tasks; our conclusions apply only to the tested regime.
\textit{(ii)}~All QUBO variants use classical simulated annealing, not quantum hardware; ``quantum-inspired'' refers to the formulation paradigm, not to quantum speedup.
\textit{(iii)}~Synthetic scaling workflows may not capture the full diversity of production HPC workloads.
\textit{(iv)}~Enhancement strategies that rely on CP-SAT guidance or HEFT repair introduce classical solver dependencies, so the observed feasibility improvements are not attributable to QUBO alone.
\textit{(v)}~The evaluation uses a fixed list-scheduling decoder; alternative decoders might shift relative orderings.

\subsection*{Applicability to HPC Practice}
For HPC practitioners, the practical implication is clear: CP-SAT and HEFT remain the recommended tools for current-scale scheduling.
QUBO formulations are best understood as a structured framework for encoding scheduling constraints in a form compatible with future quantum hardware, rather than as a replacement for classical solvers today.
The empirical regimes identified in this work (feasibility thresholds, penalty sensitivity bounds, scaling limits) provide concrete guidance for researchers evaluating quantum-inspired methods on their own scheduling instances.

\subsection*{Future Work}
Future work will extend the evaluation to larger and more diverse workflows, integrate real-world HPC traces, and explore automated penalty estimation techniques. Evaluation on emerging quantum hardware, including quantum annealers and gate-based processors, will be essential to assess whether genuine quantum advantages can be realized as hardware capabilities mature.

\section*{Acknowledgment}

The authors thank the reviewers for valuable feedback improving this work. The authors gratefully acknowledge the grant provided by NHR at NHR-Nord@G\"ottingen as part of the NHR infrastructure for presenting this research at ISC 2026 Conference, Hamburg, as well as GWDG and University of G\"ottingen for supporting to conduct this research.
All experimental data, benchmark code, and results are publicly available at \href{https://github.com/AasishKumarSharma/qubo_benchmark}{github.com/AasishKumarSharma/qubo\_benchmark} to enable reproducibility and further research.
\bibliographystyle{IEEEtran}
\bibliography{references}

\end{document}